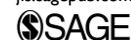

# SKOS Concepts and Natural Language Concepts: an Analysis of Latent Relationships in KOSs

## Anna Mastora
Ionian University, Faculty of Information Science & Informatics, Greece

## Manolis Peponakis
[1] Ionian University, Faculty of Information Science & Informatics, Greece
[2] National Hellenic Research Foundation / National Documentation Centre, Athens, Greece

## Sarantos Kapidakis
Ionian University, Faculty of Information Science & Informatics, Greece

## Abstract
The vehicle to represent Knowledge Organization Systems (KOSs) in the environment of the Semantic Web and linked data is the Simple Knowledge Organization System (SKOS). SKOS provides a way to assign a URI to each concept, and this URI functions as a surrogate for the concept. This fact makes of main concern the need to clarify the URIs' ontological meaning. The aim of this study is to investigate the relation between the ontological substance of KOS concepts and concepts revealed through the grammatical and syntactic formalisms of natural language. For this purpose, we examined the dividableness of concepts in specific KOSs (i.e. a thesaurus, a subject headings system and a classification scheme) by applying Natural Language Processing (NLP) techniques (i.e. morphosyntactic analysis) to the lexical representations (i.e. RDF literals) of SKOS concepts. The results of the comparative analysis reveal that, despite the use of multi-word units, thesauri tend to represent concepts in a way that can hardly be further divided conceptually, while Subject Headings and Classification Schemes – to a certain extent – comprise terms that can be decomposed into more conceptual constituents. Consequently, SKOS concepts deriving from thesauri are more likely to represent atomic conceptual units and thus be more appropriate tools for inference and reasoning. Since identifiers represent the meaning of a concept, complex concepts are neither the most appropriate nor the most efficient way of modelling a KOS for the Semantic Web.

## 1. Introduction

The Semantic Web and linked data have been receiving an increasing amount of attention in the last decade. One of their basic functions is the assignment of a Uniform Resource Identifier (URI) to each resource. In this way, a giant "machine processable" network of http-based URIs is enabled [1]. The Simple Knowledge Organization System (SKOS) is one of the fundamental pillars of the Semantic Web. It is a recommendation by W3C providing "a standard, low-cost migration path for porting existing knowledge organization systems to the Semantic Web" [2]. According to W3C, "The SKOS data model views a knowledge organization system as a concept scheme comprising a set of concepts. These SKOS concept schemes and SKOS concepts are identified by URIs, enabling anyone to refer to them unambiguously from any context, and making them a part of the World Wide Web" [2].

SKOS defines, though in a generic way, what constitutes a concept: it is perceived as "an idea or notion; a unit of thought. However, what constitutes a unit of thought is subjective, and this definition is meant to be suggestive, rather

**Corresponding author:**
Anna Mastora, Ionian University, Faculty of Information Science & Informatics, Greece

than restrictive" [2]. The basic idea behind this generic definition is the intention to provide a neutral framework for the migration of a wide diversity of Knowledge Organization Systems (KOSs) into the Semantic Web [3]. Each SKOS concept is identified by a URI and the lexical representation of the concept is a Resource Description Framework (RDF) literal. Even though the resource behind the URI may be a set of URIs, the initial URI refers to the SKOS concept as an indivisible whole. The fact that the URI is used by the software agent to refer to the ontological substance of the concept demands that the SKOS concept represents a coherent, indivisible, conceptual unit in order to have an efficient inference process. So SKOS concept URIs have come to be considered as indivisible conceptual units, which means that they cannot be broken to smaller constituents in an inference process. It is this characteristic of SKOS URIs which brings them to the attention of this study.

Taking a step back, it is acknowledged that this literal is actually the label of a KOS concept, which mainly is expressed in natural language, albeit somewhat restricted. However, natural language is both ambiguous and has grammatical and syntactic characteristics which denote relations, either conjunctive or disjunctive. Then, a natural presumption is that a KOS concept may contain multiple conceptual units which makes disputable the semantics of the derivatives of the inference process because the component parts, although easily perceivable as constituents by the human brain, are transformed into an indivisible conceptual unit for the computer. Since inference is performed through URIs (in the context of the Semantic Web), the question, then, is how these distinctive natural language concepts are embedded and represented in the structures of the Semantic Web.

Therefore, the nature of these constituents, as well as their relations, has to be investigated so as to designate the atomic concepts, i.e. the conceptually undividable units, by analyzing KOS concepts on the basis of their linguistic characteristics. According to Giunchiglia et al, an "atomic concept is a class, an entity, a quality or value in the domain" [4] and their identification (through natural language) plays an important role during the modelling process [5]. Their designation is very important in the context of the Semantic Web because the atomic level is the node, i.e. a URI, and "One node represents one object" [6]. Ultimately, discovering whether a KOS concept can be decomposed into smaller constituents would contribute both to an assessment of the suitability of its migration to a SKOS concept and designate potential applications in KOS editing in order for it to be more suitable for SKOS in the context of the Semantic Web.

The rest of the study is organized as follows: Section 2 presents the background analysis of the field consisting of information about SKOS, the representation of concepts within certain types of KOSs as well as information about natural language in terms of grammatical and syntactic rules. Section 3 consists of the aim and objectives of this study. Section 4 comprises the scope and the methodology followed. In section 5 the results are presented and in section 6, we discuss issues that arose during the realization of the study, which leads to section 7, where our conclusions are drawn.

## 2. Background analysis

In this study, there are three main areas of interest. First is SKOS and how it represents KOSs in the Semantic Web, particularly the use of URIs for representing KOS concepts, and the implications of URIs' use for inference. The second area is the notion of concept within KOSs viewed from the natural language perspective. Finally, there is the natural language itself with its grammatical and syntactic rules and its potential for analysis through computational linguistics.

### 2.1. Simple Knowledge Organization System (SKOS)

SKOS is a recommendation by W3C for the representation of structured controlled vocabularies such as thesauri, classification schemes, and subject headings systems and aims at the representation of data in order to become machine processable in the environment of the semantic web and linked data. As already mentioned, the SKOS data model views a knowledge organization system as a concept scheme comprising a set of concepts and these SKOS concepts are identified by URIs [2]. For the lexical representation of the concept an RDF literal is used. These literals could contain a code to declare the language of the text; for example `"text"@en` in standard Turtle syntax, or `<rdfs:label xml:lang="en">` in RDFS XML coding. Properties like `skos:prefLabel`, `skos:altLabel` and `skos:hiddenLabel` are used to correlate the URI of the concept with its lexical representations. By using SKOS it is not possible to declare relationships between the different appellations of the same concept. To model the appellations, like declaring that label B stands as an acronym for label A, the "SKOSeXtensionforLabels" should be used [7]. In the context of the semantic web the hierarchy of SKOS concepts and thus the inference or reasoning for software agents is mainly performed by building on SKOS URIs and not on RDF literals, i.e. the lexical representations of these URIs. However, under certain circumstances the values of literals could be used for inference, as in the cases of dates or predefined values (e.g. language, in Figure 1). Yet, the basic mechanism of inference is based upon the URIs.

Thus, the point of interest is the ontological substance of concepts and not their names. These names (lexical representations) could be used in querying or as descriptive information which explains to the user the meaning of the underlying concepts in a human understandable language. Figure 1 visualizes a fragment of a thesaurus (Eurovoc) and the relation ("`skos:exactMatch`") of a specific concept with another thesaurus (GEMET), all expressed in SKOS. The oval shapes represent nodes which are identified by a URI and they could serve as objects or subjects to new triples. The rectangle shapes represent literals which are end-points and they cannot be subject to new connections.

As already mentioned, although it does not provide a restrictive definition about what a concept is, SKOS defines sufficiently and restrictively the relations between concepts. It "distinguishes between two basic categories of semantic relation: **hierarchical** and **associative**. A hierarchical link between two concepts indicates that one is in some way more general ("broader") than the other ("narrower"). An associative link between two concepts indicates that the two are inherently "related", but that one is **not** in any way more general than the other" [2]. The hierarchical relationships could be a rather generic subsumption relationship of broader-narrower or they could be considered as transitive, i.e. functioning as an "is-a" relationship. For example, if the concept "Canary" has a broader concept of "Bird" and a narrower concept of "Frilled canary", these relations could be specified as `skos:broaderTransitive` and `skos:narrowerTransitive` respectively. This transitivity allows an agent to assume that every "Frilled canary" is a "Bird". On the contrary, in the case of the concept "Vehicles" ([http://id.loc.gov/authorities/subjects/sh85142531](http://id.loc.gov/authorities/subjects/sh85142531)) in LC authorities, for which the broader concept is "Transportation" and the narrower concept, "Wheels", the aforementioned inference is not valid because, obviously, every "Wheel" is not a means of "Transportation". At this point it should be stated that such clarity and consistency in the definition of relationships is rare in traditional KOSs, therefore the use of the SKOS transitive property is equally rare, as shown in the SKOS Implementation Report [8].

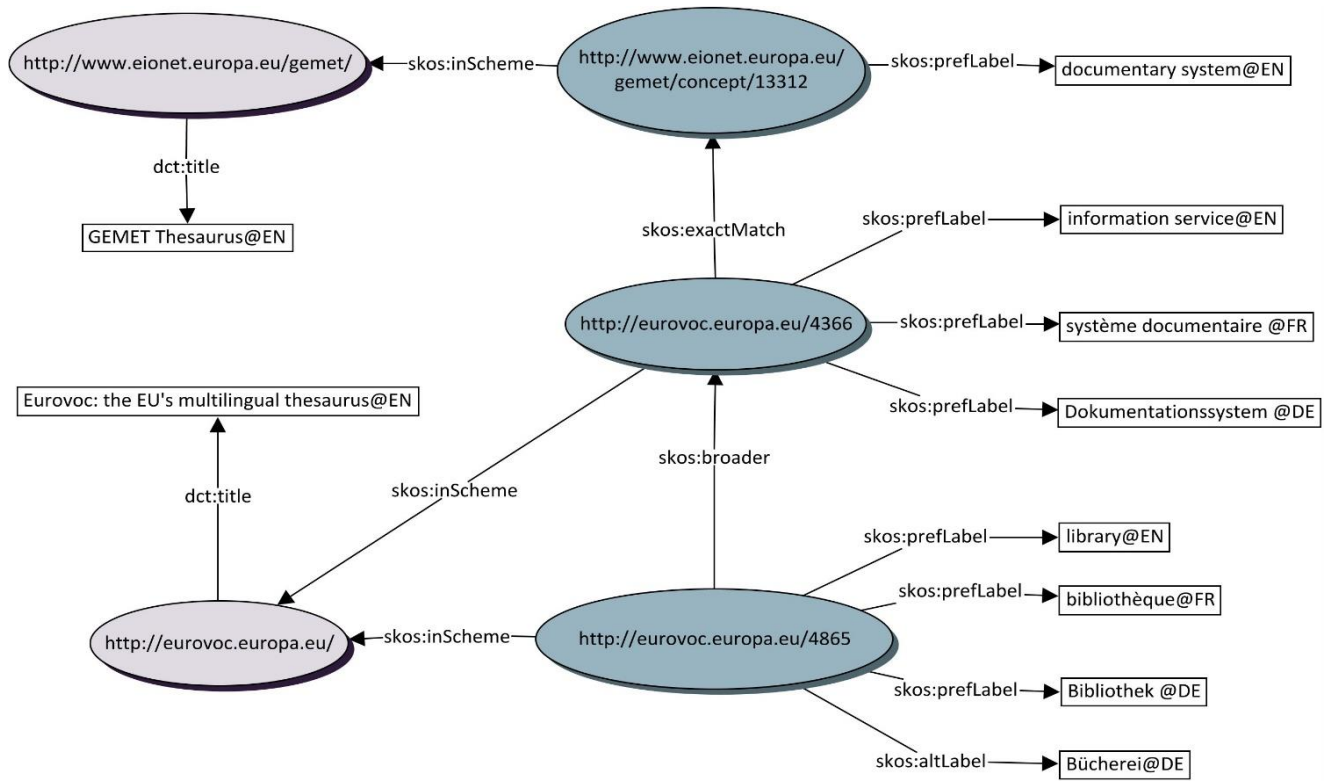

**Figure 1**.Graphical representation of three concepts and their relationships from two different thesauri expressed in SKOS. The labels of the concepts from Eurovoc appear in various languages while the one from GEMET only in English.

For an effective use of SKOS relations, KOS terms must comprise atomic concepts. When they do not, inferred relations are ambiguous or controversial. For instance, the Dewey Decimal Classification (DDC) term "747.3" stands for "Ceiling, walls, doors, windows" which is a subcategory of "Interior decoration". As a SKOS URI, this term is used as a single concept, without one being able to refer to either "doors" or "windows" individually. Therefore, every attempt to semantically relate any of the constituents of the "747.3" term (concept) with another concept, consequentially, applies the same semantic relation to all its constituents. But what would be the semantic outcome if "Ceiling, walls, doors, windows" was related to a specific material like "bricks" with the predicate "made of"?

## 2.2. Concepts in KOS and Natural Language

Knowledge Organization Systems (KOSs) "are mechanisms for organizing information" [9]. They are artefacts which "should be understood as systems basically organizing concepts and their semantic relations" [10] and they "are used to organize materials for the purpose of retrieval and to manage a collection" [9]. Basically, a KOS term mainly stands for the concept this term represents, as delivered in any KOS through time and space. "Concept-based IR represents both documents and queries using semantic concepts, instead of (or in addition to) keywords, and performs retrieval in that concept space" [11]. KOSs are meant to organize information, either for a particular domain or with a broader coverage, serving, in any case and in principle, as a convention for expressing an area of interest in a way that its constituents are effectively communicated towards and among the targeted audience. And, depending on their expressiveness, KOSs are also structured on the basis of various kinds of explicit relationships, such as hierarchical, equivalence (synonymy), relatedness and translation.

In this paper we abide by the principle that "concepts should be considered the building blocks of all forms of KOS" [12]. A priori, in the KOSs' universe, an atomic concept is each term of the KOS. But what is the correlation of the KOS concept with natural language concepts, so that each of them can represent the other with consistency? This is fundamental since the Semantic Web is oriented to concept-centric approaches. SKOS itself "has implied the predominance of the *concept-based thesaurus* approach over the *term-based thesaurus* approach" [13]. But each type of KOS defines in its own context and through its own rules what constitutes a concept. For the purpose of this study, we refer to three representative KOSs from three different categories: a thesaurus (Eurovoc) [14], a subject headings system (Library of Congress Subject Headings - LCSH) [15] and a classification scheme (Dewey Decimal Classification - DDC) [16]. All are available in SKOS.

A Thesaurus is defined by ISO 25964 as a "controlled and structured vocabulary in which concepts are represented by terms, organized so that relationships between concepts are made explicit, and preferred terms are accompanied by lead-in entries for synonyms or quasi-synonyms" [17]. Each term can be one word or a compound term which is a term "that can be split morphologically into separate components" [17]. Traditionally, thesauri are lexical resources carrying semantic information on instances of a certain natural language, mainly expressing this information through synonymy relations. A Subject Headings system is a controlled vocabulary which also uses terms in a way similar to thesauri. Additionally, these terms may be subdivided by the addition of a subheading. LCSH was one of the first libraries' vocabularies which was converted into SKOS [18]. Classification schemes have different origins. The principle 4.14 in the DDC introduction declares that: "Since the parts of the DDC are arranged by discipline, not subject, a subject may appear in more than one class". DDC is primarily expressed in the form of notations but, according to Mitchel [19], the notation (i.e. DDC number), the lexical representation of the notation and the URI of a DDC class may be used interchangeably as long as they represent the same subject.

Thus, for all the above KOSs, the SKOS URI is the indivisible unit for the representation of a concept into the semantic web. This principle leads to the problem area of how indivisible concepts are treated at language level. In terms of syntax, we accept here that the fundamental unit of natural language is the word, which Pinker calls the "syntactic atom" [20] in a sense that it cannot be split. It is a fact, though, that there is no one-to-one relation between words and concepts. And although a noun often stands for a concept, in many cases a concept is represented by a multi-word unit which may comprise various parts-of-speech, like "Social Anthropology" (adjective, noun) or "Mainframe Computers" (noun, noun). Additionally, there are cases where the collocation of the words forms a whole new meaning which is different from the meaning of each of the parameter-words, as in the case of "greenhouse effect". The aforementioned cases are considered compound terms; therefore they stand for atomic concepts as perceived in the natural language context.

On the other hand, the usage of composite terms is common practice in KOSs. These composite terms comprise two or more natural language concepts with a kind of relation between them. For example, the heading "Sports for children" (LCSH http://id.loc.gov/authorities/subjects/sh85126925) could not be considered as an atomic concept, since it could be represented as two different KOS terms, i.e. "Sports" and "Children" with a specific predicate to relate them. At this point an additional remark should be made. Actually, in the example "Sports for children", these concepts are also expressed independently within the same KOS, i.e. LCSH. As there is no provision for formalistically declaring this specific kind of relationship between these two concepts, a natural language relation is used, i.e. "for", and a composite term is formed instead. However, even if there was such a provision on the KOS part, this type of relation could not be brought to SKOS.

Other cases of divisible KOS terms comprise the ones which enumerate natural language concepts, either with the use of commas (asyndeton schema) or with conjunctions (and, or). In these cases there clearly is no dependency relation; therefore, the assumption that these formations are independent conceptual units is mostly true.

Furthermore, at a different level, in many cases there is no one-to-one correlation between languages, not even in the case of scientific terminology. It is then that lexical gaps emerge, in a sense that certain concepts "do not have a succinct representation in a given language. However, they can be expressed as a free combination of words" [21]. So, in any given language, anything that can be expressed as a single noun may be expressed as a multiword unit in another language. For example, the German word "Zeitgeist" has no analogue in English; instead the multiword unit "spirit of the age" could describe the meaning.

The URIs of the semantic web are sufficient for efficient processing of information by software agents but "if the Semantic Web is to be queried by humans, there will be no other way than dealing with the ambiguousness of human language" [22] because users will consult the URIs and they will understand their meaning mainly through their labels, i.e. RDF literals. So, there is the need for a "mechanism that maps natural language texts to these concepts" [11]. A representative example of such ambiguousness is the following: Eurovoc comprises the term "*underground literature*" as a non-preferred term of "*grey literature*" classified under the Broader Term "*document*". In LCSH the term "*underground literature*" is classified under "*literature*" and the scope note explains "*here are entered works about publications issued clandestinely and contrary to government regulation*".

## 2.3. Natural Language: morphology, syntax and computational linguistics

Natural language is both fundamental and complicated as a communication system; therefore it has been the subject of many disciplines. Linguists, psychologists, neuroscientists, philosophers and computer scientists have all contributed to the studying of natural language. For the purpose of this study the focus lies with the linguistic perspective and particularly with the morphological and syntactic aspect in terms of meaning extraction. Rightfully, this is a study area of semantics. And since it has been established that words and meaning are not in a one-to-one relationship, the quest of meaning can take many routes. One of them is defined by Portner who thoroughly discusses the ideas of formal semantics through various constructions including grammatical and syntactic formations. He states that "most famously supported by the philosopher Quine, the theory of [meaning] holism claims that the meaning of a word or phrase or sentence depends on its relationships with other words, phrases, and sentences" [23]. It is this functional aspect of linguistic units which is the particular subject of our study.

Natural language has rules, norms and patterns concerning its morphology and syntax. Some rules are strict; some others are subject to a looser implementation. The morphology and the austerity of rules differ among various natural languages. "Linguists group the words of a language into classes (sets) which show similar syntactic behaviour, and often a typical semantic type. These word classes are otherwise called syntactic or *grammatical categories,* but more commonly still by the traditional name *parts* of speech (POS)" [24]. Concerning the morphology of language, a Part Of Speech denotes its use either autonomously or within a certain context. At the syntax level, a Part Of Speech is seen with regards to its position within a phrase in relation to the other Parts Of Speech in the same phrase. Murphy [25] stated that "syntactic variables (like the presence or absence of an article), and morphological variables (like verb vs. noun endings on words) can be used as cues about what the word may mean". In his work on semantic primitives, Wilks states that "A primitive (or rather a set of primitives plus a syntax etc.) is a reduction device which yields a semantic representation for a natural language via a translation algorithm and which is not plausibly explicated in terms of or reducible to other entities of the same type" [26]. He goes on to discuss an example about a 'noun' which, in linguistics, is not further reducible and is not normally considered to be explicable in terms of entities of any other type. But it is not only nouns or one-word terms that can be considered as non-dividable units. Sometimes more nouns are required to form a non-dividable conceptual unit, as already mentioned in the previous section.

In all languages, words (and entire word classes) can be divided into the two broad classes of content (or major class) words and function (or minor class) words. Nouns, verbs, adjectives, and adverbs are content words, and adpositions, conjunctions, and articles, as well as auxiliaries and words classified as `particles' are function words [27]. Although this is a generally acceptable classification, for the purpose of this study, some kinds of adverbs, like *today*, *now* and *more*, are considered as function words. Content word classes are generally open (i.e., they accept new members in principle) and large (comprising hundreds or thousands of words), and content words tend to have a specific, concrete meaning. They tend to be fairly long (often disyllabic or longer), and their text frequency is fairly low. By contrast, function word classes are generally closed, i.e. made of finite sets and hardly expandable. Function words tend to have abstract, general meaning (or no meaning at all, but only a grammatical function in specific constructions, i.e. concerning the correlations of neighbouring terms). They tend to be quite short (rarely longer than a syllable), and their text frequency is high. The high frequency and the cases where function words act like modifiers make them a worthy area of study, particularly within KOSs in regard to meaning extraction.

After verbs and prepositions, conjunctions are the most important carriers of semantic relations in (complex) sentences. While verbs and prepositions typically determine the inner structure of elementary situations, conjunctions are the carriers of the semantic relations between different situations [28]. This fact is of major importance when sparse data is studied, as is the case with the three KOSs we are studying, since they mostly comprise elliptical sentences.

In his work, Booij stated that "the relationship between inflectional forms and their semantics is not that straightforward. Thus, inflectional phenomena give a perfect illustration of how complex the relation between form and meaning in natural languages can be" [29]. Nevertheless, he reported some morphological markings on nouns, verbs and adjectives that are mostly common in the languages of the world. For nouns, some markings are the Number (singular, plural, dual etc), the Case (nominative, genitive, accusative etc.), Definiteness and Gender, and for Adjectives, Degree (positive, comparative, superlative), and again Number, Gender, Case and Definiteness. He also identified common markings for verbs such as Tense (present, past, future), Aspect (imperfective, perfective etc), Mood (indicative, subjunctive, imperative etc.), Voice (active, passive etc.), Number (singular, plural etc), Person (first, second, third) and Gender. These findings are also supported by Murphy [25] who stated that "It is our basic concepts of objects, event, and properties that lead languages to tend to have similar morphosyntactic properties." Additionally, it would be both interesting and informative for the reader to browse The World Atlas of Language Structures [30] for a thorough presentation of language structures and use it as a tool for comparing certain features among the languages of the world. It is evident that there are features which commonly apply in a number of languages, independently of the family they belong to. For example, Greek, which is a highly inflectional language, shares the "*order of noun and genitive*" with Dutch, French, German and Hebrew, while the *noun-phrase conjunction* of *"and* being different from *with"* is shared with English, Finnish, Hebrew and Russian. Greek also shares the "*adjective-noun order"* with Dutch, English, Finnish, German and Turkish, while the *periphrastic causative construction*, which is classified as "*purposive but non-sequential",* is common in Greek, Finnish, Hebrew, Korean, Maori and Russian, among others.

"The field of computational linguistics (CL), together with its engineering domain of natural language processing (NLP), has exploded in recent years. The focus of research in CL and NLP has shifted over the past three decades from the study of small prototypes and theoretical models to robust learning and processing systems applied to large corpora." [31]. "Most NLP systems today are not monolithic entities, but rather consist of distinct components, often arranged in a processing pipeline. For example, identification of semantic role (e.g., agent, patient, theme) depends on syntactic parsing, which in turn depends on part-of-speech tagging, which in turn depends on tokenization"[ 31]. In this study we are particularly interested in the potential of the *parsing* process which is referred to as "the process of automatically analyzing a given sentence, viewed as a sequence of words, in order to determine its possible underlying syntactic structures" [31]. This process is of interest with regards to the semantics it can convey through formalization.

## 3. Aim and objectives

The aim of the study is to demonstrate the extent to which the label of SKOS concepts – as expressed in KOS terms – can/should decompose into smaller constituents towards assuming better results for the inference process. For this purpose, the study examined the natural language description of labels originating from representative KOSs, i.e. a thesaurus, a subject headings system, and a classification scheme. In order to fulfil the aim of the study, the following questions were addressed.

- What is the natural language evidence which advocates towards the divisibility of KOS terms?
- What is the natural language evidence which advocates towards the indivisibility of KOS terms?
- To what extent is each type of evidence present in the selected KOSs?
- To what extent is each KOS appropriate for migration to SKOS?

The study attempts to answer the aforementioned questions by conducting morphosyntactic analysis, i.e. exploit the typological classification of languages in terms of their morphological and syntactic characteristics, of the RDF literals (i.e. the labels) of SKOS concepts by implementing Natural Language Processing techniques, i.e. Part-of-Speech (POS) tagging. Through morphosyntactic analysis, the latent intra-term (not inter-term) relationships are designated, which denote the presence of more than one natural-language concept within the KOS terms.

Briefly, for the purpose of this study, we used three KOSs, a POS tagger and a suite of tools for managing and querying the data in order to group the data based on certain criteria. More information on the scope and methodology is given in the following section.

## 4. Methodology

In this section, information is deployed on the methodology we followed, as well as on the tools used, along with certain decisions and restrictions that apply for the management and analysis of the data. In order to address the questions of this study three types of KOSs were selected, namely the multilingual thesaurus Eurovoc [14], the Library of Congress Subject Headings (LCSH) [15], and the Dewey Decimal Classification (DDC) system [16]. Thesauri, subject headings and classification schemes are the three most frequent kinds of KOSs used in the context of libraries.

At this point we have to mention that we used the Greek versions for examining the literals. Such a choice was made due to two contradictory facts and a counterbalancing solution. Firstly, KOSs comprise elliptical phrases, that is, mainly noun phrases with few words, which means that they offer limited contextual information in terms of morphosyntactic analysis. Secondly, POS taggers need rich contextual information in order to efficiently assign the right types of words during the morphosyntactic analysis. In the studied literals, with an average of 2.93 words per entry (ranging from one word to a maximum of 25 words per entry), the POS tagger would face challenges in determining the right types of words. For example, in English, the word "turn" isolated from a richer context could stand either for a verb or a noun. The counterbalance is Greek's rich morphology, i.e. high inflectionality, which overcomes to a certain extent the aforementioned limitations due to the fact that it is more likely to define accurately and unconditionally a part of speech, regardless of its syntactic context. This attribute makes Greek more appropriate for our analysis. Still, we have to note that in Greek, too, there are some cases of morphosyntactic ambiguousness. One such case is encountered in the forms of the plural feminine of nouns in nominative or accusative case and the second person of verbs in singular number, as in the case of "προσθέσεις" (prosthéseis), which stands for either "additions" or (you) "add".

The Eurovoc thesaurus is multidisciplinary and available in Greek. The version used was reproduced and adapted from the original language editions of the *Eurovoc Thesaurus (Edition 4.3)* ©European Communities, 2008 (http://Eurovoc.europa.eu/) and contained 6,882 descriptors. In terms of LCSH, the May 2013 complete authorities version by the National Library of Greece (NLG) was used, which contains the Greek translation of selected LC Subject Headings, and is referred to in this text as *LSCH (NLG corpus)*. From 55,204 subject headings in total we isolated the ones that were declared to be a translation of LCSH, numbering 19,986 records. The entries we finally used were at the subfield level, amounting to 30,073 overall. After having performed subfield deduplication, the uniquely identified subfields reached 10,308. We chose to work in the subfield level because we consider that subfields constitute the smallest autonomous functional units of a heading. Finally, concerning DDC, we used the 13th abridged edition translated into Greek by the Greek National Documentation Centre. DDC delivered 5,398 entries, which were reduced to 3,811 by the deduplication process.

Before proceeding any further, we have to define the terms "entries", "tokens" and "words", as used in the context of this study. An *entry* is each sequence of characters or groups of characters which is defined as an individual listing within the contents of KOSs. More specifically, an entry for Eurovoc is each preferred term, i.e. descriptor; for LSCH (NLG corpus) it is every subfield; and for DDC it is the lexical representation of each decimal notation. Notes of any kind are not included as entries. In this study the words "entries" and "terms" – when referring to KOSs in general – are used interchangeably, but a distinction was considered appropriate for presenting the metrics so that there would be no confusion with the word "terms" and its established use in the context of thesauri. *Tokens* are every significant unit (character or group of characters) which was assigned a POS characterization; it may be a word, digit(s) or punctuation mark. *Words* are the remaining sequences of characters if we exclude digits and all punctuation marks as well as other non-lexical symbols from the identified tokens.

The morphosyntactic analysis was implemented using the POS tagger developed by the Greek Institute for Language and Speech Processing (ILSP) [32]. The ILSP-NLP POS tagger handles Greek texts and is freely available as a web service at http://nlp.ilsp.gr/soaplab2-axis/. In the case of our study, the input was an xml document containing the KOS entries (one entry per row) and the output was an xml document with the POS tagged entries and the lemmas for each token. Figure 2 depicts an excerpt of the tagger's output. The Eurovoc entry is the term "χάρτης εκπαιδευτικών ιδρυμάτων" (chártes ekpaideutikón idrymátōn), which stands for "map [of] educational institutions". The outcome is the morphosyntactic annotation for each word, along with its respective lemma.

The full tag set is available at http://nlp.ilsp.gr/nlp/tagset_examples/tagset_en/. In this study we only present the identified-within-our-data parts of speech. It should be noted that some pre-processing of data was necessary to remove punctuation that caused problems, such as the % and * symbols. Additionally, some post-processing was also required so that the data would be computable.

```
Eurovoc entry:
<p id="EUR.224">χάρτης εκπαιδευτικών ιδρυμάτων</p>
Tagger's output:
 <p id="EUR.224">
  <s id="s224" casing="lowercase">
    <t id="t544" word="χάρτης" tag="NoCmMaSgNm" lemma="χάρτης"/>
    <t id="t545" word="εκπαιδευτικών" tag="AjBaNePlGe" lemma="εκπαιδευτικός"/>
    <t id="t546" word="ιδρυμάτων" tag="NoCmNePlGe" lemma="ίδρυμα"/>
  </s>
 </p>
```

**Figure 2.** Excerpt of tagger's output.

The direct outcome of the tagger is only part of the required solution. The POS tagging process already described has provided the clues towards forming the criteria for classifying the KOS entries (terms) as divisible or indivisible with regards to the conceptual units they carry. To give an example of the analysis process, let us consider the term "Oil and natural gas". The morphosyntactic analysis reveals the Parts-of-Speech, i.e. two Nouns (oil, gas), one Adjective (natural) and one Conjunction (and). The fact alone that there are two nouns in this elliptical (multiword) phrase alerts about the existence of multiple concepts, since most concepts are expressed by nouns. But it takes more to draw a definite conclusion. It is the co-occurrences of parts-of-speech along with their morphosyntactic characteristics that will lead to more accurate answers. In the above example, the analysis reveals that the four words comprise two natural language concepts, i.e. "oil" and "natural gas". This assumption is strengthened by the syntactic pattern (Noun - Conjunction - Adjective - Noun) which implies that the conjunction actually connects (apparently with no dependency relationship) the two nouns, one of which is accompanied by an adjective. The noun and the adjective form one conceptual unit and the other noun is another conceptual unit. The use of the conjunction "and" has brought the two concepts under one KOS term. We need to clarify, though, that this example is a simplification, matching the low inflectionality of the English language. In highly inflected languages more complex structures need to be analyzed, as in the case of Greek.

Therefore, so far, for delivering the results of the study regarding the typology of the Greek language, the outcome of the POS tagging process and the identified POS co-occurrences, i.e. syntactic patterns, are examined for each KOS. The purpose of this workflow is to designate the criteria for classifying the groups of labels for the ten most frequent syntactic patterns for each KOS into either divisible or indivisible conceptual units. In the results' analysis section the reason for choosing the ten most commonly occurring patterns is justified. Furthermore, the study examines a number of criteria in particular (effective for the overall set of data, not just the ten most common patterns) which advocate towards the divisibility of the literals and are presented individually in the results' analysis.

In addition to the above criteria, we categorized the parts-of-speech on the basis of "open" or "closed" word classes, as they are described and defined in subsection 2.3. In this study there is particular interest in the words of the "closed" word class, among which, *prepositions* have a more significant role since they can transform the dependencies between syntactic units, whereas *articles* have little effect in that regard. *Pronouns* may serve syntactically as nouns, yet they offer no additional information without their assignment to the word they substitute. Therefore, their designation in the context of a phrase may convey meaning as long as they are assigned to certain words. Clearly, the "open" word class carries the opposite features, with their main characteristic being that they are potentially infinite. Such words are *nouns*, *verbs* or *adjectives*. Generally, *adverbs* may be considered either closed or open according to whether they derive from *adjectives* or not. *Adverbs* that derive from *adjectives* are grouped in the open word class inheriting their "openness" from the *adjectives* class. Here, we consider *adverbs* as an "open" word class due to the fact that additions to the overall group are possible.

The notation "inflected" – "invariable" refers to the discrimination of Parts of Speech according to whether or not, in terms of inflections, they can take various grammatical forms.

## 5. Results Analysis

In this section we present the metrics and their analysis resulting from the Part-of-Speech tagging procedure and the outcome of the KOS terms categorization according to the criteria already referred to. The specific criteria are deployed later in this section. Table 1 depicts some general purpose rates concerning the recorded number of entries, words and tokens, as well as the average number of words and tokens per entry for each of the examined KOSs.

As already mentioned in the Methodology section, the first outcome of this study was the distribution of the designated Parts of Speech in each KOS as shown in Table 2; rates refer to the number of tokens per KOS. As depicted, and with regards to the total number of words for each KOS, Eurovoc is comprised of 4.10% of "closed" class words, LCSH (NLG corpus) of 8.42%, and DDC of 19.15%. Overall, the number of POS with a rather significant presence is very limited. Excluding *common nouns* and *adjectives*, in Eurovoc only five (four if punctuation is also excluded) POS exceed the rate of 1%, while seven (or five without punctuation) in LCSH (NLG corpus), and nine (or seven without punctuation) in DDC. It was also observed that although verbs – along with nouns – are normally considered as the head of a clause, in the case of KOSs there are very few verbs detected. This means that, consequently, the functional characteristics of verbs, which mainly denote action, are absent in KOSs. It is made clear then that KOSs almost exclusively store nominal phrases, which are considered elliptical (short) sentences. As shown in Table 2, *nouns*, and specifically *common nouns*, are the most frequent POS in all three KOSs.

**Table 1.** KOS entries, tokens and words.

|                          | Eurovoc | LCSH (NLG corpus) | DDC    |
|--------------------------|---------|-------------------|--------|
| Number of entries        | 6,882   | 10,308            | 3,811  |
| Number of tokens         | 15,234  | 23,670            | 16,414 |
| Number of words          | 15,067  | 20,497            | 14,613 |
| Words per entry (avg)    | 2.19    | 1.99              | 3.83   |
| Tokens per entry (avg)   | 2.21    | 2.30              | 4.31   |

Also worth mentioning are the rates for *Proper nouns*, which, in terms of tokens per KOS, comprise 4.29% in Eurovoc, 5.11% in LCSH (NLG corpus) and 6.07% in DDC. The inclusion of Proper Nouns within the KOSs is part of a broader discussion on whether a Named Entity is itself considered to be a concept or is an instance of a generic concept. It appears that to a certain extent some Named Entities (here in the form of *Proper Nouns*) are considered by KOSs as concepts.

But, even though it is generally accepted that nouns are important meaning carriers, Table 2 cannot provide all the required information for the attempted analysis. It mostly offers clues which have to be further analyzed considering the information recorded in Table 3, which lists, in descending order, the ten most frequent co-occurrences of POS combinations for each analyzed KOS. Therefore, the analysis also involves the designation of POS co-occurrences in order to classify the conceptual units as either divisible or indivisible.

Additional information, such as the grammatical case of words, is also sometimes required for carrying out the aim of this study. Let us assume the case of Dewey's entry *Mésa tekmēríōsēs, ekpaideutiká mésa, mesa enēmérōsēs dēmosiographía; ekdotikḗ* (Μέσα τεκμηρίωσης, εκπαιδευτικά μέσα, μέσα ενημέρωσης· δημοσιογραφία· εκδοτική), which stands for "*Documentary media, educational media, news media*" (DDC 070.1). We isolate the word "mésa" for further analysis. The word "mésa" is translated as "in" or "within" as an adverb or "media" as a noun. In this example, it would be incorrect to tag the word as an adverb. Taking into account the fact that in the case of "ekpaideutiká mésa" there is a preceding adjective (neuter, plural, nominative) then the word "mésa" is a (neuter, plural, nominative) noun. This decision is also assisted by the position of the word which is at the end of the elliptical phrase (*ekpaideutiká mésa*) which would be the right place for an adverb if it accompanied a verbal, not nominal, phrase. In the case of "*Mésa tekmēríōsēs*", again, it would be incorrect to define the word "mesa" as an adverb, since it accompanies a noun in the genitive case, i.e. the structure is [Noun Nominative + Noun Genitive], which is a common way for expressing dependency or specialization of a certain entity. The position of the word, i.e. at the beginning of the phrase, is, again, an indication of it being a noun and not an adverb. Instead, in the case of "Metakínēsē plēthysmỗn pros, apó, mésa se koinótētes" (Μετακίνηση πληθυσμών προς, από, μέσα σε κοινότητες), which stands for "Movement of people to, from, within communities" (DDC 307.2), it would be accurate to designate the word "mésa" as an adverb since it is followed by the preposition "to" and a noun, which makes it a common case of an adverbial prepositional phrase.

The job of unambiguously designating the parts of speech is, mainly, the job of the POS tagger. But, it is also part of this study to take into consideration such rules in order to define the levels and the criteria of analysis. Therefore, assigning a word to a certain part of speech alone does not say much about the actual meaning of the word. For this purpose, apart from the general purpose data already mentioned, Tables 3 and 4 present the additional information needed for Table 5, part of the core results of the study.

**Table 2.** Part Of Speech (POS) tagging of KOSs.

| Part-of-Speech (POS) | POS type | EUROVOC | LCSH (NLG corpus) | DDC |
|---|---|---|---|---|
| Adjectives [open], [inflected] | Basic | 3678 (24.14%) | 4929 (20.82%) | 3179 (19.37%) |
| | Comparative | 7 (0.05%) | 18 (0.08%) | 82 (0.50%) |
| | Superlative | 9 (0.06%) | 4 (0.02%) | 1 (0.01%) |
| Articles [closed],[inflected] | Definite | 101 (0.66%) | 337 (1.42%) | 843 (5.14%) |
| | Indefinite | 3 (0.02%) | 0 (0.00%) | 2 (0.01%) |
| Nouns [open], [inflected] | Common | 8443 (55.42%) | 11825 (49.96%) | 6443 (39.25%) |
| | Proper | 653 (4.29%) | 1209 (5.11%) | 997 (6.07%) |
| Pronouns [closed], [inflected] | Demonstrative | 0 (0.00%) | 0 (0.00%) | 8 (0.05%) |
| | Indefinite | 0 (0.00%) | 5 (0.02%) | 98 (0.60%) |
| | Personal | 0 (0.00%) | 0 (0.00%) | 4 (0.02%) |
| | Possessive | 5 (0.03%) | 27 (0.11%) | 12 (0.07%) |
| | Relative | 6 (0.04%) | 0 (0.00%) | 36 (0.22%) |
| | Relative indefinite | 2 (0.01%) | 0 (0.00%) | 0 (0.00%) |
| Verbs [open], [inflected] | Indicative | 18 (0.12%) | 34 (0.14%) | 166 (1.01%) |
| | Participle | 27 (0.18%) | 21 (0.09%) | 80 (0.49%) |
| Numerals [open], [inflected] | Cardinal | 8 (0.05%) | 19 (0.08%) | 4 (0.02%) |
| | Ordinal | 15 (0.10%) | 18 (0.08%) | 23 (0.14%) |
| | Multiplicative | 5 (0.03%) | 1 (0.00%) | 1 (0.01%) |
| Adverbs [open], [invariable] | Basic | 82 (0.54%) | 213 (0.90%) | 220 (1.34%) |
| | Comparative | 1 (0.01%) | 0 (0.00%) | 0 (0.00%) |
| | Superlative | 0 (0.00%) | 0 (0.00%) | 0 (0.00%) |
| Adpositions [closed], [invariable] | Prepart | 69 (0.45%) | 539 (2.28%) | 195 (1.19%) |
| | Simple | 299 (1.96%) | 187 (0.79%) | 601 (3.66%) |
| Conjunctions [closed], [invariable] | Coordinative | 98 (0.64%) | 614 (2.59%) | 970 (5.91%) |
| Particles [closed], [invariable] | Negative | 35 (0.23%) | 17 (0.07%) | 30 (0.18%) |
| Residual [open], [n/a] | Foreign word | 314 (2.06%) | 276 (1.17%) | 95 (0.58%) |
| | Symbol | 0 (0.00%) | 4 (0.02%) | 0 (0.00%) |
| Abbreviations [open], [n/a] | All CAPS | 240 (1.58%) | 11 (0.05%) | 5 (0.03%) |
| | Others | 0 (0.00%) | 61 (0.26%) | 45 (0.27%) |
| Punctuation [n/a], [n/a] | Comma | 2 (0.01%) | 1495 (6.32%) | 946 (5.76%) |
| | Terminal | 0 (0.00%) | 2 (0.01%) | 30 (0.18%) |
| | All others | 203 (1.33%) | 1691 (7.14%) | 910 (5.54%) |
| Digits [n/a], [n/a] | Numbers etc | 2 (0.01%) | 110 (0.46%) | 388 (2.36%) |

A worthy observation of the data in Table 3 is about the overall proportion of the first ten syntactic patterns in each KOS. In Eurovoc and in LCSH (NLG corpus), they cover more or less 80% of the overall identified syntactic patterns, while in DDC they cover almost 44% of the overall patterns. Related information not depicted in Table 3 is that there are many unique syntactic patterns in each KOS. Eurovoc has 149 combinations appearing only once out of a total of 286 unique syntactic patterns; LCSH (NLG corpus) has 266 syntactic patterns appearing only once out of 474 unique patterns; and, DDC has 855 combinations appearing only once out of 1,128 patterns. The above data advocates towards the decision of including the ten most frequently occurring syntactic patterns in the core of the analysis. Still, although certain syntactic patterns carry enough information for implementing the aforementioned classification, some of the patterns need deeper linguistic analysis, beyond the POS level. Therefore, in order to provide answers regarding the questions of this study about divisibility or indivisibility rates among the labels, each of the first ten syntactic patterns (that is, seventeen unique patterns) were examined in terms of the Greek language typology with regards to respective criteria, as presented below.

**Table 3.** Ten most frequent syntactic patterns (POS co-occurrences) in KOSs and divisibility characterization.

| Eurovoc | | LCSH (NLG corpus) | | DDC | |
|---|---|---|---|---|---|
| Syntactic Pattern [and divisibility] | Frequency | Syntactic Pattern [and divisibility] | Frequency | Syntactic Pattern [and divisibility] | Frequency |
| Adj+N [i] | 2,182 (31.71%) | N [i] | 3,353 (32.53%) | N [i] | 634 (16.64%) |
| N [i] | 1,364 (19.82%) | Adj+N [i] | 1,936 (18.78%) | Adj+N [i] | 387 (10.15%) |
| N+N [f.a.] | 819 (11.90%) | N+Punct+Adj [i] | 677 (6.57%) | N+N [f.a.] | 156 (4.09%) |
| N+Art+N [f.a.] | 454 (6.60%) | N+N [f.a.] | 548 (5.32%) | N+Conj+N [d] | 112 (2.94%) |
| Res [i] | 191 (2.78%) | N(N) [d] | 369 (3.58%) | Dig [i] | 112 (2.94%) |
| N+Adj+N [f.a.] | 171 (2.48%) | N+Adp+N [d] | 361 (3.50%) | Adj [i] | 63 (1.65%) |
| Adj+Adj+N [i] | 143 (2.08%) | N+Conj+N [d] | 311 (3.02%) | N+Adj+N [f.a.] | 61 (1.60%) |
| N+Adp+N [d] | 118 (1.71%) | Adj [i] | 261 (2.53%) | Adj+Adj+N [i] | 50 (1.31%) |
| Adj [i] | 103 (1.50%) | Adj+Adj [i] | 134 (1.30%) | N+Art+N [f.a.] | 44 (1.15%) |
| Adj+N+N [f.a.] | 93 (1.35%) | Adj+N+Punct+Adj [i] | 124 (1.20%) | Adj+N+Conj+N [d] | 40 (1.05%) |
| Sum | 5,638 (81.92%) | Sum | 8,084 (78.33%) | Sum | 1,659 (43.53%) |

In Table 3 the abbreviations, in alphabetical order, are explained as follows: Adj = Adjective // Adp = Adposition // Art = Article // Conj = Conjunction // CPunct = Close Punctuation // Dig = Digits // N = Noun // OPunct = Open Punctuation // Punct = Punctuation // Res = Residual (Foreign Word) // V = Verb. Concerning the information in square brackets: [i] is for "indivisible term", [d] is for "divisible term" and [f.a.] declares that this syntactic pattern was subject to "further analysis" in order to decide upon its divisibility.

## 5.1. Syntactic pattern analysis in terms of indivisibility

The criteria which emerged from our analysis are largely common to the ones set in the work of Gavriilidou and Lambropoulou [33] who used a hybrid method for term extraction from Greek corpora, combining both a qualitative (morphosyntactic or shallow parsing) and quantitative (statistical) approach. First, the criteria designating the indivisibility of terms are presented.

### 5.1.1. One-word terms

As depicted in Table 3, the single noun pattern (e.g. *athletes, education, ceremonies*) is one of the most common in all three KOSs. In Eurovoc it is the second most common pattern [1,364 (19.82%)], while in LCSH (NLG corpus) it is the most common [3,353 (32.53%)], as it is for DDC [634 (16.64%)]. Within the first ten syntactic patterns, the rates of single-word terms (not just nouns) which apparently comprise coherent conceptual units are 1,658 (24.09 %) in Eurovoc, 3,614 (35.06 %) in LCSH (NLG corpus) and 809 (21.23 %) in DDC, while the overall rates of single-word terms, also including cases not depicted in Table 3, are 1,687 (24.51 %) in Eurovoc, 3,660 (35.51 %) in LCSH (NLG corpus) and 837 (21.96 %) in DDC.

### 5.1.2. Adjective + Noun (Adj + N)

One way to specialize a reference is to use adjectives, of which there is a high proportion among the parts of speech within the three KOSs; these are very important for our analysis. "Murphy and Andrew […] asked subjects to provide synonyms or antonyms for adjectives. The words, however, were presented as parts of phrases, such as fresh water, fresh air, fresh bread, and so on. The results showed that the antonyms and synonyms supplied changed depending on the other word in the phrase. For example, a synonym for fresh water might be spring water, yet clearly spring bread is not a synonym for fresh bread. The adjective fresh seems to take on somewhat more specific meanings depending on the noun it modifies. The specific meanings led subjects to provide different synonyms in different contexts" [25]. The use of adjectives implies that they further define the words they accompany, for example *Technological advances*. Especially when found within terms with very few words (mostly up to three), it is rather safe to claim that they are parts of indivisible terms. Such cases refer to either attributive or nominal adjectives, for example *Urban architecture*. Therefore, the rates of adjectives alone are not sufficient to fully analyze their contribution in identifying the semantic constituents which this study is seeking. This is why adjectives are seen in more details when considering the syntactic patterns in which they are found. The analysis shows that the [Adj + N] pattern in Eurovoc comes first and involves 31.71% of the entries, against 18.78% in LCSH (NLG) and 10.15%) in DDC. More syntactic patterns including adjectives are discussed later.

As it happens, all three KOSs share the first two most frequent POS co-occurrences, i.e. [Adj + N] and [N], although the rates of DDC are significantly low. Additionally, in all three KOSs the inverted form of the first combination, i.e. [N

+ Adj], is also present but with significantly less frequency (32, 45 and 19 times respectively), and this is why they are not included in Table 3.

### 5.1.3. Noun + Noun (N + N)

Another popular tag combination is the [N + N]. A significant observation about this pattern concerns the grammatical cases of the nouns involved. Of particular interest is the pattern [Noun in Nominative (NoNm) + Noun in Genitive (NoGe)], such as *ασφάλεια ζωής* (aspháleia zōḗs, standing for *life insurance* in Greek but with an inverted word order) since the genitive case is the basic grammatical case for denoting dependency. The comparison and interpretation of the statistics presented in Table 4 are quite intriguing. The first row shows the instances where a Noun in the Nominative case is followed by a Noun in the Genitive. In addition to the use of the genitive case, the terms (which are also considered indivisible) comprise two nouns in the nominative (for example *αδελφές ψυχές* (adelphés psyches, meaning *soul-mates*) which in Greek is represented by two individual words) or one noun in the nominative and another in the accusative. In Eurovoc, as shown in the second row of Table 4, there are 95 (1.38%) terms with such characteristics, in LCSH (NLG corpus) there are 241 (2.34%) and in DDC 50 (1.31%). The third row depicts the sum of these two cases, while the last row depicts the number of the overall rates of the [N + N] syntactic pattern (also depicted in Table 3). As it happens, all [N + N] syntactic patterns in Eurovoc are included in the two specific categories designating indivisible terms and clearly most cases involve the use of the Genitive. In LCSH (NLG corpus) there is a minor deviation of two, while in DDC there are eight cases not complying with the rule.

**Table 4.** Further analysis of the Noun – Noun pattern.

| | Eurovoc | LCSH (NLG corpus) | DDC |
|---|---|---|---|
| [No + No] with 2nd No in Ge | 724 *(10.52%)* | 305 *(2.96%)* | 98 *(2.57 %)* |
| [No + No] with Nm - Ac combination | 95 *(1.38%)* | 241 *(2.34%)* | 50 *(1.31%)* |
| SUBSUM 1st & 2nd subsets | 819 | 546 | 148 |
| SUM [No + No] syntactic pattern | 819 *(11.90%)* | 548 *(5.32%)* | 156 *(4.09%)* |

No=Noun, Nm=Nominative, Ge=Genitive, Ac=Accusative

In general, all two-word terms consist mostly of a head and a qualifier which are, therefore, used together. Consequently, all two-word formations within the examined KOSs are considered indivisible terms under the prerequisite that no punctuation or other symbol is contained. For concluding with the two-word terms, we have to mention at this point that the recorded syntactic pattern [Adj + Adj] raised some questions during the analysis, since two adjectives could not actually form a meaningful entry. Further examination of the specific cases [appearing 134 times (1.30%) only in LCSH (NLG corpus)] proved that they are actually nouns which the POS tagger characterised wrongfully in this particular environment but, morphologically, they could be adjectives. Nevertheless, they all fall into one of the patterns [Adj/N + N/Adj] discussed in this section; therefore, the overall rates of indivisible terms are not compromised. The same situation occurred with the single adjective pattern discussed earlier in the section of one-word terms. But again, further analysis revealed that the recorded cases comprise one of the following: nouns (mainly female nouns in singular number representing abstract concepts, like Cybernetics) or nominalised adjectives. In this case, too, overall rates of indivisible terms falling in the category of one-word, irrespective of their subcategory, are valid. In addition, it should be mentioned that adjectives appearing in other syntactic patterns are correctly tagged since the POS tagger had more contextual information and performed better.

### 5.1.4. Noun + comma + Noun/Adjective (N + comma + N/Adj)

Tag combinations of the general syntactic form *word, word* are found in LCSH (NLG corpus) 898 times (8.71%), while DDC has only one entry (0.03%) of the *Noun, Noun* pattern. This pattern appears in 120 entries (1.74%) in LSCH (NLG corpus); while the *Noun, Adjective* syntactic pattern appears in 677 entries (9.84%). Eurovoc has no such case. The terms of this form denote some kind of inversion, such as "Literature, Portuguese"; therefore they are considered as coherent conceptual units, i.e. indivisible terms.

### 5.1.5. Adjective + Adjective + Noun (Adj + Adj + N)

This pattern, which is one of the common syntactic patterns denoting indivisible terms, such as *ακαθάριστο εθνικό προϊόν* (*akatháristo ethnikó proïón*, i.e. *gross national product*), appears in 143 (2.08%) entries in Eurovoc and in 50 (1.31%) in DDC. In LCSH (NLG corpus) there are 48 entries (0.47%), which is below the threshold of the first ten syntactic patterns; this is why it is not included in Table 3. In the typology of the Greek language, such formations comprise indivisible conceptual units since the adjectives are the qualifiers of the noun.

### 5.1.6. Adjective + Noun + Noun (Adj + N + N)

This pattern appears in the first ten most frequently occurring syntactic patterns only in Eurovoc, i.e. 93 entries *(1.35%)*. In LCSH (NLG corpus) the pattern appears in 35 entries *(0.34%)* and in DDC in 21 entries *(0.55%)*. In order for this pattern to be considered an indivisible conceptual unit, two conditions should apply: either the second noun should be in the Genitive or all words should be in the same case, for example *διακυβερνητική νομική πράξη* (*diakybernētikḗ nomikḗ práxē*, i.e. *intergovernmental legal instrument*). These conditions apply in 92 out of 93 cases in Eurovoc, 34 out of 35 times in LCSH (NLG corpus) and in all 21 entries in DDC.

### 5.1.7. Noun + Adjective + Noun (N + Adj + N)

This pattern appears within the first ten in Eurovoc (171 entries, *2.48%*) and in DDC (61 entries, *1.60%*). In order for this pattern to comprise an indivisible conceptual unit, the adjective and the second noun have to be in the Genitive, for example *αλλαγή πολιτικού καθεστώτος* (allagḗ politikoú kathestṓtos, i.e. *change of political system*) or *μέσο μαγνητικής εγγραφής* (*méso magnētikḗs engraphḗs*, which stands for *magnetic medium*). All the recorded cases within the first ten syntactic patterns are of this form.

### 5.1.8. Noun + Article + Noun (N + Art + N)

This pattern  comprises terms such as *στρατιωτικοποίηση του διαστήματος* (*stratiōtikopoíēsē tou diastḗmatos*, i.e. *militarization of space*), which means that in order for this pattern to carry an indivisible term, the article and the second noun has to be in the Genitive. All recorded cases of this pattern follow the rule. Eurovoc has 454 entries (*6.60%*) and DDC has 44 (*1.15%*). LCSH (NLG corpus) also has 108 entries (*1.05%*), but they do not appear among the ten most frequent.

### 5.1.9. Adjective + Noun + punctuation + Adjective (Adj + N + Punct + Adj)

This type of pattern, which implies the presence of an indivisible term, is met in 124 entries (*1.20%*) only in the LCSH (NLG corpus) and responds to terms of the form *Χριστιανική Λογοτεχνία, Βυζαντινή* (*Christianikḗ logotechnía, Byzantinḗ*, i.e. *Christian literature, Byzantine*).

## 5.2. Criteria denoting divisibility

Following the criteria for indivisibility, the criteria and the respective results of divisibility are presented below. It should be noted here that the divisibility criteria are implemented in the overall data, not just the first ten syntactic patterns, but the four which have remained undiscussed are covered by these latter criteria. In addition, the criteria for divisibility form two subcategories. One is "enumeration/parataxis", and involves the first two criteria, i.e. *conjunctions* and *commas*, which recite concepts without declaring any kind of relationship between them. And second, "composites" which comprise at least two concepts related with either an adposition (e.g. literature for children) or with the use of parentheses.

### 5.2.1. Conjunctions

Concerning grammar, conjunctions are "and" as well as "or". This category also contains expressions denoting the meaning of "et cetera" in any possible lexical representation, like "etc", "and so forth", or "et al". The hermeneutics of conjunctions include both Linguistics and Logic [30], which makes them a rather significant category as regards meaning extraction. In Table 2 the total number of conjunctions is presented for each KOS. For a deeper analysis, more evidence is provided here. Eurovoc has 98 (1.42%) "and-conjunctions", while no "or-conjunction" is designated. Since sometimes more than one conjunction appears within an entry, the number of entries (terms) that carry conjunctions was also recorded. In Eurovoc only one term contains more conjunctions. In LCSH (NLG corpus) there are 611 (5.93%) "and-conjunctions" and 3 (0.03%) "or-conjunctions". The number of entries containing conjunctions is equal to the recorded number of conjunctions, but it should be mentioned that the three cases with "or-conjunctions" also contain "and-conjunctions". Finally, DDC has 958 (25.14%) "and-conjunctions" and 12 (0.31%) "or-conjunctions". The former derive from 899 entries, while the latter are as many as the entries. Since, as in LCSH (NLG corpus), three cases with "or-conjunctions" also contain "and-conjunctions", the number of entries containing at least one conjunction is 908 (23.83%).

As already mentioned, this category also contains expressions denoting the meaning of "et cetera" in any possible lexical representation, such as "etc", "and so forth", or "et al". In Eurovoc, though, there is no such case, while in DDC there are 16 (0.42%) entries of the form [and + indefinite pronoun (other)]. In LCSH (NLG) there are 53 (0.51%) entries mainly consisting of different formations of the abbreviation "etc". Only two are of the form [and + indefinite pronoun (other)]. The entries with the indefinite pronoun are already recorded during the initial metrics (the ones about "and-conjunctions"), while the 52 (0.5%) entries of LCSH (NLG), which comprise abbreviations, should be calculated in addition to the initial number.

### 5.2.2. Commas

The use of commas, particularly in cases where more than one comma is involved, were of interest due to the fact that this pattern implies the presence of the *asyndeton schema*, or *parataxis*, which is the conjunction of individual elements with no dependency relations. While in Table 2 the total number of commas is depicted, we located and counted the number of entries where commas are used and discovered that Eurovoc contains only two entries with commas (0.03%), LCSH (NLG corpus) contains 1,426 (13.83%) and, finally, DDC contains 459 entries (12.04%). Further analysis reveals that in LCSH (NLG corpus) the number of terms containing two or more commas are 53 (3.72%) out of 1,426 and in DDC 378 (82.35%) out of 459. Cases where only one comma was detected were discussed earlier in the context of the syntactic pattern *Noun, Noun/Adj*.

### 5.2.3. Parentheses

The analysis of the data revealed that the use of parentheses denotes the presence of individual conceptual units. In Eurovoc there are 95 entries with parentheses, in LCSH (NLG corpus) 837 entries and in DDC, 380. Parentheses are used in various occasions. One common combination is the form *Noun (Noun)* which appears in 369 LCSH (NLG corpus) entries (3.58%), while DDC has only 11 (0.29%).

### 5.2.4. Adpositions

As already mentioned, *Adpositions* comprise words like *from*, *for*, *in*, *to*, *by*, *based*, *on*, *against*, *due* and *with*. Although greatly ignored in data processing, they play an important role because, even though they do not carry meaning in their own right, they define semantically the words they accompany, e.g. "Offenses against the person". In Table 2 the overall number of the designated adpositions is depicted, i.e. 368 in Eurovoc, 726 in LCSH (NLG corpus) and 796 in DDC. The number of entries carrying at least one adposition is as follows: 357 (5.19%) in Eurovoc, 714 (6.93%) in LCSH (NLG corpus) and 612 (16.06%) in DDC.

## 5.3. Remarks on indivisibility and divisibility results

In section 5.1 the criteria and the rates about the indivisibility of terms are covered concerning the ten most frequent syntactic patterns in all three KOSs. A more detailed presentation would include more patterns but there are at least two reasons that these are not included in this study. Firstly, the plethora of the syntactic patterns which is already described in the initial description of Table 3, especially in DDC, prohibits the formation of collective rules. Each pattern would demand individual presentation. Secondly, the rates seem rather insignificant with regard to the coverage of the first ten patterns. The proportion of terms examined reveals a repeated motive among the KOS terms which allows for the characterization as indivisible for more than 80% of the Eurovoc terms, 68.21% of LCSH (NLG corpus), but only for 39.33% of the DDC terms.

**Table 5**. Distribution and types of divisible terms per KOS.

| | Eurovoc | LCSH (NLG corpus) | DDC |
|---|---|---|---|
| Type: Enumeration/ parataxis | 97 (1.41%) | 704 (6.83%) | 1224 (32.12%) |
| Type: Composite | 449 (6.52%) | 1504 (14.59%) | 894 (23.46%) |
| SUM (unique) | 536 (7.79%) | 2118 (20.55%) | 1766 (46.34%) |

On the other hand, in section 5.2 the criteria denoting divisibility are presented accompanied with respective frequencies and rates. Table 5 depicts the categorization (along with respective percentages) of patterns that imply the presence of divisible terms with regards to the criteria and results already discussed in section 5.2.

## 6. Discussion

The conversion of traditional data to RDF is not a sufficient condition to acquire the full potential of the Semantic Web [34]. Even more, in cases such as the traditional KOSs, where there is a lack of "well-defined semantics and structural consistency", some kind of "reengineering" is necessary [35]. On the other hand new sorts of modelling tend to encourage humans to think of new ways for knowledge representation: for example, the ontological formalisms of OWL force ontology modellers to produce literals which do not exist in the specialized dictionaries of a certain domain's vocabulary [36]. If we want efficient machine-processable concepts, then it is necessary for them to obtain a solid ontological substance. If the point is to build on concepts for inference and reasoning, all relations among them must be explicitly declared in a formalistic way, and this cannot be implemented only at the structural level. As long as KOSs treat concepts as complex units with latent relations, the potential remains limited. Therefore, the more non-formalistic, latent relations exist within an entry, the less effective is its computational processing, because, even if the entry could/should be divided into sub-units, there is no way for the computer, i.e. there are no explicit rules, to perform such a division. Instead, once indivisible conceptual units are declared, synthesis of information can be performed. Especially in the context of SKOS, where the emphasis is placed on the URI and not on the lexical representation, the logical indivisibility of a concept is vital. On the other hand, we could not overlook the fact that a synthesis process requires the implementation of more complicated and sophisticated algorithms that require more computational power which, in some cases, may prove prohibitive for effective, real-world implementations [37].

The terminological aspect of KOSs is not particularly addressed in this study because it is considered rather inherent and generic. It is not within the scope of this study to question the theories and practices of terminology; we rather accept that the terms appearing in KOSs are commonly accepted expressions (at least in the context of the Kuhnian Paradigm) for the concepts they have been designated. Therefore, when we refer to the SKOS concept, a URI is assigned to each concept and the agents use this URI to refer to the ontological substance of the concept. In this case the notion of ontological substance refers to the fact that the concept is not constrained by its lexical expression. As shown in figure 1; GEMET's "documentary system" is the same concept (`skos:exactMatch`) as Eurovoc's "information system". However, the ontological substance is not related to cognition processes of the human mind and the Wittgensteinian open texture, nor to the change which is performed on the perception of a certain concept through time, nor is it based on variant epistemological theories [38]. It is not within the scope of this study to examine whether the concept "Europe" refers to its current countries or to an era that there were no countries at all. This distinction is one that may or may not be adopted by a KOS. If the KOS provides various perspectives for "Europe", then a respective number of URIs is expected. Therefore, it has to be clear that the relationship between epistemology and LIS concerning the "conceptual stability and change" [10] of concepts is out of the scope of this study. This was a deliberate decision taken on the grounds that in the environment of the Semantic Web the notion of meaning is not strictly attached to understanding, since "It is not necessary for intelligent agents to understand information; it is sufficient for them to process information effectively, which sometimes causes people to think the machine really understands" [1]. Hence, if it is a necessary for the concept "Europe" to express different epistemological aspects in the context of SKOS, then, as a necessary and sufficient condition, there should be as many URIs as there are perspectives.

"Linked Data supplies users with comprehensive knowledge, but also raises a need to search entities in Linked Data. Entity search aims to find entities based on a query" [39]. So there is the need to move from the ontological substance of the entity to its names and vice versa. Consequently, it is critical to develop mechanisms which perform these transitions with accuracy. These mechanisms cannot be developed without studying language itself and its own mechanisms since it, i.e. natural language, will be the intermediate in the human-computer interaction. For example, collocations are an important issue when it comes to information extraction. A collocation is a sequence of consecutive words which carry the properties of a syntactic and semantic unit, the meaning of which is not derived by the meaning of its parameter words. Such formations cause difficulties because their syntactic pattern or even the literals themselves cannot be considered apart from their semantics.

The fact that the algorithms of natural language processing tools are trained, ideally, in large corpora containing proper sentences (and not elliptical) is perhaps the reason for their poor performance when it comes to analyzing sparse data. In our study, the POS tagger performed better in the DDC analysis, which has the highest rate of words per entry (3.83) as well as the highest rate of maximum words per entry (max=25); meanwhile, in the analysis of the LCSH (NLG corpus) data, which has the lowest rate of words per entry (1.99) and the lowest rate of maximum words per entry (max=8), the POS tagger faced severe challenges. The POS tagging in Eurovoc was of average performance (avg 2.21 words per entry), for which the highest number of words per entry is 13.

A representative case of the challenges faced by the POS tagger is the observation of entries which do not contain a *head of clause* part of speech. This can be justified to some extent by studying the data qualitatively, since some of the justified cases comprise of abbreviations, digits or another logical no-noun word combination, e.g. [Adj + Conj + Adj], as in the case of "good and bad". By studying these cases, which admittedly are not that many, we are able to assign them to POS tagging errors; or, better said, to POS tagging weaknesses in correctly identifying the parts of speech without enough contextual information. The same questions arise in other cases such as in the presence of one or more adverbs.

Stop-words are usually ignored during either pre-processing or post-processing of any text corpora with the excuse that they often appear within documents or that they hold no semantic information. We argue that the decision to ignore stop-words is not a decision that fits all purposes. What is defined as a stop-word does carry semantic meaning in the context of multi-word units when they represent a single concept. Also, such words may indicate a specific type of relation between two natural language concepts. For example in the case of LCSH SKOS concept "Films by children" (URI http://id.loc.gov/authorities/subjects/sh85048240) the word "by" is vital for the interpretation of the meaning. Generally speaking, the utilization and exploitation of "closed" word classes is recommended since it can differentiate circumjacent words and assist towards disambiguating collocated word classes since syntactic dependencies tend to appear in a specific arrangement within a specific language.

## 7. Conclusions

KOS concepts are expressed to a great extent in multi-word units. These multi-word units are elliptical phrases which are analyzed based on the typology of language, and more specifically upon morpho-syntactic characteristics concerning the designated parts of speech, according to whether they represent one or more concepts. This study has shown that an atomic concept is often represented by multi-word units containing more than one noun (for example "mutual assistance scheme" in Eurovoc). The examination of the ten most frequent POS patterns among the examined KOSs reveals a repeated motive among the KOS terms which allows for the characterization as indivisible for more than 80% of the Eurovoc terms, 68.21% of LCSH (NLG corpus), but only for 39.33% of the DDC terms.

On the other hand, there is a remarkable ratio of KOS terms which can be decomposed to multiple conceptual components based on linguistic criteria and evidence. In this study, the divisible terms were grouped into two categories. One involves the enumeration (parataxis) of concepts, for example the term "Dogmatism, eclecticism, liberalism, syncretism, traditionalism" in DDC. The other category is composite terms, such as "Women on bank notes" in LCSH. As depicted in Table 5, parataxis is very common in DDC since it involves almost 1/3 of the terms, while it is less common in LCSH (NLG corpus) and rare in Eurovoc. DDC also incorporates many composite terms (23.46%), which are also met in LCSH (NLG corpus) (14.59%), as well as in Eurovoc (6.52%).

In all the studied KOSs there was no full correspondence between what constitutes a concept in natural language and what constitutes a concept in KOS. Significantly closer to concepts as they are expressed through natural language concepts are the terms of the thesaurus, while the classification scheme concepts are more distant. A variety of natural language concept and relation types underlies what is considered by each of the KOSs to be an indivisible unit. If the indivisible building blocks of KOSs contain multiple conceptual units, then the potential of expressiveness is limited. Conversely, the more broken down to its fundamental components the semantic information is, the more chances we have for a detailed and accurate conceptual analysis.

Given the fact that there is a need for the use of atomic concepts in the context of the Semantic Web, the ratio of non-atomic concepts per KOS indicates its suitability for being expressed in SKOS. Our study presents evidence towards making this judgment. Therefore, in addition to the critique of Panzer and Zeng [40], our study suggests that DDC is not suitable for migration to SKOS due to the extensive use of non-atomic concepts, especially the phenomenon of parataxis. Concerning the subject headings, they are somewhat appropriate but not entirely so. On the contrary, the thesaurus, as supported by evidence, is far more appropriate for this kind of migration. Eventually, this evidence shows that some re-engineering of KOSs is necessary in order to fulfil their new purpose in the context of the Semantic Web.

Thus the concepts of traditional KOSs, as they migrate into the semantic web through SKOS URIs, carry information which does not correspond to the smallest piece of information (atomic concept in our context). For as long as natural language has been the basic mechanism for inference and reasoning this has not been a big problem. But now that identifiers (i.e. URIs), either as concepts or as relations between concepts, are being used to represent, in an indivisible way, the meaning of a concept, such complex concepts are neither the most appropriate nor the most efficient way of modelling a KOS for the Semantic Web.

## Funding

This research has been partially co-financed by the European Union (European Social Fund – ESF) and Greek national funds through the Operational Program "Education and Lifelong Learning" of the National Strategic Reference Framework (NSRF) - Research Funding Program: Heracleitus II.